# May 11-12 Extreme Space Weather Events Brief and Dose Rate Model Response

**V. A. Burov[1], K. I. Kholodkov[1,2] (0000-0003-0324-9795), I. M. Aleshin[1,2,3] (0000-0001-7489-2951)**

[1]Fedorov Institute of Applied Geophysics, Moscow, Russia
[2]Schmidt Institute of Physics of the Earth of the Russian Academy of Sciences, Moscow, Russia
[3]Geophysical Center of the Russian Academy of Sciences, Moscow, Russia
* **Correspondence to:** Kirill Kholodkov, keir@ifz.ru.

**Abstract:** In this brief paper, we analyze space weather events that occurred on May 11 and 12, 2024, from the perspective of an operational space weather center that provides advisories for civil aviation. One of the key metrics monitored by the center is the radiation dose rate at operational flight altitudes. A model implemented by the center provides the dose rate in real time. The model showed that dangerous levels were momentarily exceeded just above the usual 30,000 feet level during the events. This paper highlights differences in models used by various space weather centers, emphasizing the need for harmonization.

Space weather is a set of phenomena in interplanetary space that occurs as a result of changes in the Sun. Over the past decade, there has been a steady increase in society's awareness of the fact that space weather represents a substantial threat to the technological infrastructure[1,2,3,4,5]. In some publications [6], the impact of space weather phenomena has been explored, with a particular emphasis on the response of the ionosphere in relation to coronal mass ejections (CMEs) as the primary hazard. Recently, several scholars have directed their attention specifically towards the impact [7] on positioning accuracy during periods of ionospheric disturbances. Other studies have shown the effects of powerful X-ray flares on the ionosphere.

Today, aviation relies heavily on technologies that are vulnerable to space weather disturbances. Prominent examples of these technologies are global navigation satellite systems (GNSS) and over-the-horizon high-frequency (HF) radio communications. X-ray flares can cause serious problems for precision positioning and GNSS (Global Navigation Satellite System) navigation services. Solar flares have been shown to affect navigation services for up to several hours, leading to critical situations in various navigation applications [2, 6]. These energetic protons have the potential to reach Earth and pose a threat to aircraft operating in polar regions by degrading HF-communication capabilities.

However, it's not only technology that suffers from space weather disturbances. Aircraft flying at typical commercial and corporate airline altitudes are constantly exposed to high-energy charged particles and secondary neutrons of cosmic origin. These types of particles, known as galactic cosmic rays (GCR) that originate outside our solar system, and solar energetic particles (SEPs) can affect aircraft microelectronics systems and the health of airline crew members and passengers [8,9]. Flights on high-latitude or intercontinental routes may exceed the maximum public and fetal exposure limits during a single solar energetic particle event and through multiple (~5-10) high-latitude round-trip flights due to GCR exposure. It is important to note that while some countries monitor airline crew radiation exposure, many do not, leaving airline crews as the only occupational group exposed to unquantified and undocumented radiation levels over their careers.

The discussion about the significant impact of space weather on aviation has increased since the beginning of the 21st century. Following the IATA's letter to ICAO in November 2011 requesting a discussion on space weather's impact on aviation, ICAO has been evaluating the use of space weather data in civil aviation. The discussion found its way into the amendment to Annex 3 of Meteorological Service for International Air Navigation. The document regulates the form and way the information on the space weather phenomena reaches the civil aviation stakeholders. The informational message is called the advisory and comes in moderate and severe form. The thresholds for moderate and severe advisories are fixed in the same document. ICAO also initiated a process to establish space weather centers in 2017. Twenty-two countries expressed interest in becoming space weather information providers. Finally, three groups were selected: the US, PECASUS (consisting of Finland, the UK, Germany, Poland, Austria, Italy, the Netherlands, Belgium, and Cyprus), and ACFJ (comprising Australia, Canada, France, and Japan), which have been designated as global centers by ICAO. As of 2022, CRC (China-Russia Consortium) joined as another global center.

China-Russia Consortium [10] consists of three organisations: Fedorov Institute of Applied Geophysics (IAG) of Roshydromet, Aviation Meteorological Center (AMC) of Civil Aviation Administration of China and National Center for Space Weather of China Meteorological Administration. Currently, IAG and AMC both act as full-featured space weather centers in round-robin, backing up each other. Besides the duty, CRC organisations also perform research and analysis tasks to improve methods, instruments and regulations. IAG has performed an analysis of recent extreme space weather events in terms of potential industry response. The event is special in a way that it highlighted the differences in methods used to compute the effect of the particular space weather phenomena.

The extreme space weather events observed on May 11-12 were caused by the passage of active region #3664. Hereby we will use the NOAA active region classification. This was the most intense group in this cycle of solar activity, with an area reaching 2,400 millionths of the Sun's visible hemisphere, 20 times the size of the Earth. The magnetic configuration was complex, with a beta-gamma-delta pattern, and there were about 50 multipolar spots, large electric currents. In addition, the maximum activity occurred during its passage across the solar disk, where the position of the group was optimal for impacting the Earth's magnetosphere. During this time, there were 6 X-ray class flares, some accompanied by solar particle events, which caused significant difficulties in radio communication and navigation. These events were further amplified by the strongest magnetic storm in the past 21 years. This complex of phenomena resulted in the extreme space weather conditions. Disturbances that are classified as extreme manifestations of solar activity, similar to the "Halloween" storms that occurred in October-November 2003, have been observed.

On May 9, two X-class solar flares were observed within this group. The first flare, X2.2, occurred at 09:13 UT, and the second, X1.1, occurred at 09:17:44 UT. During May 8 and 9, four CME events were recorded, all of which were classified as geoeffective. It is important to note that only those CME events that are directed towards Earth can have a geoeffect, and these events constitute a minority. Three of the CMEs were expected to arrive at Earth on May 10 at around 10:00 UT +/- 10 hours.

On May 10, three more flares occurred (see Fig. 1), with the first being X3.9 at 16:40 UT and the others being X5.8 and X1.5 at 18:01:23 UT and 19:43:33 UT respectively. The proton flux from these flares was recorded at a peak flux value of 207 PFUs (Particle Flux Units, particles × cm-1 × s-1 × sr-1) at 17:40 UT. This event was classified as a solar proton event.The event began at 01:40 UT following the X5 flare and the X3 flash. Both of these events were linked to the coronal mass ejection (CME).

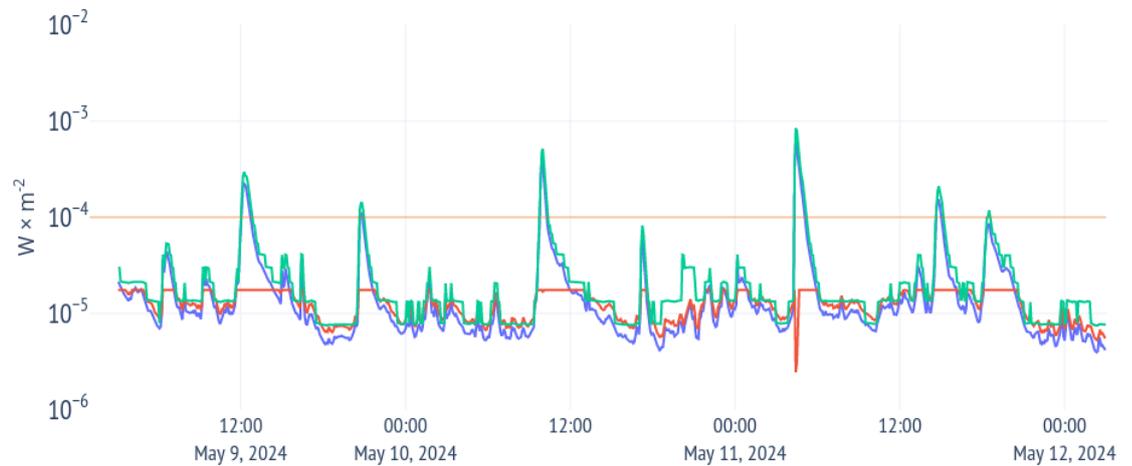

**Fig. 1.** X-Ray flux (in watts × m-2) at GEO from the GOES-16 spacecraft (blue) and GOMS-5 (Elektro-L N4) mission (red - precise measurements, green - coarse measurements). Orange line is the X1 threshold.

On May 10, the speed of the solar wind near Earth doubled to approximately 700 km/s after the arrival of the CME. On May 11, it reached a maximum value of 993 km/s. The peak total intensity of the interplanetary magnetic field (IMF) was 56 nT, and the range of the north-south component (Bz) was between +22 and -50 nT. During this period, Bz was predominantly oriented southward.

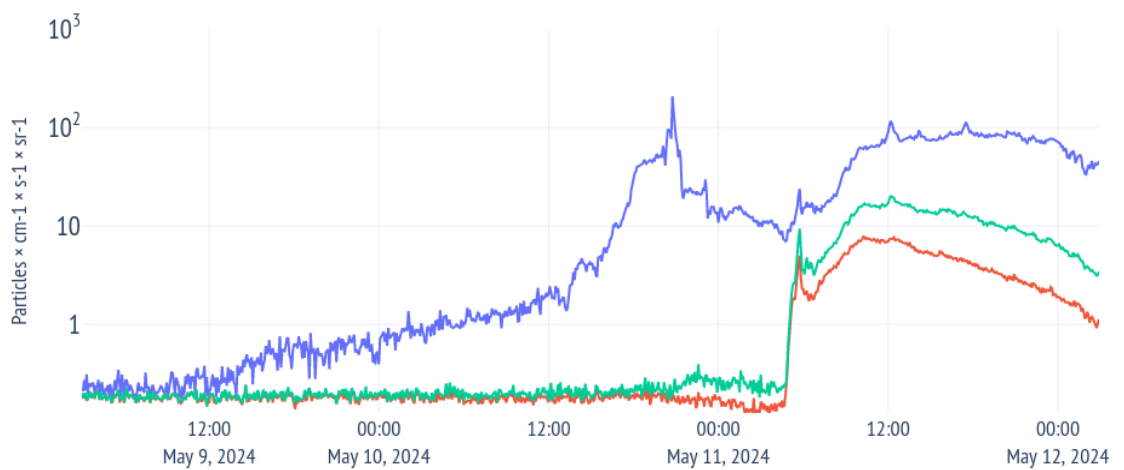

**Fig. 2.** Proton flux (in PFU) at GEO of GOES-series spacecraft. Protons with energies >=10 MeV in blue, >=100 MeV in red, >=50 MeV in green.

Proton fluxes observed during this time were relatively low (level S2) as seen on Fig. 2. However, an important aspect of this event is worth noting: a geomagnetic storm occurred (see K-indices on Fig. 3). The extent to which proton fluxes from solar flares penetrate and generate secondary particles that create cosmic radiation is dependent not only on the initial proton flux density, but also on the disruption of the magnetic field. Large geomagnetic storms can cause increased penetration and higher dose rates in the atmosphere compared to when there are no disturbances. Our dose calculation model takes this factor into account, as well as variations in the spectrum of primary radiation fluxes.

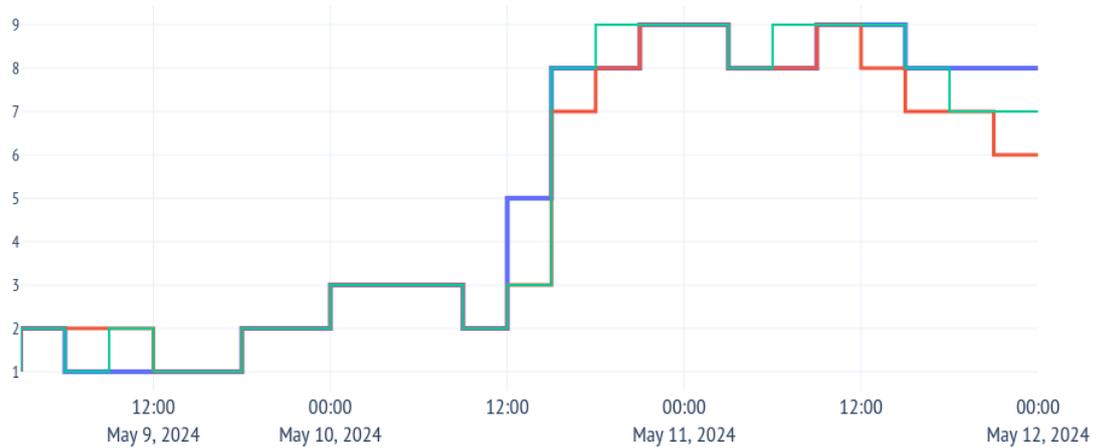

**Fig. 3**. Planetary K-indexes from GFZ German Research Centre for Geosciences (blue), Institute of Applied Geophysics (red), and Space Weather Prediction Center (green).

According to our model, if CRC was operating (on-duty) at the time, we would have issued a moderate type advisory on May 11 between 03 and 08 UTC. Fig. 4 shows the examples of dose rate maps with contours of radiation dose rate in mSv/hour for an altitude of 12.2 km (40,000 ft).

The on-duty center (ACFJ) issued advisories on communication and positioning degradation but no radiation dose rate advisory. Every space weather center operates their own radiation dose rate model. According to the model that was used by the on-duty ACFJ during the period this event did not require the advisory to be issued. We believe that the differences in our estimates of the solar flare dose rate are due to different methods of calculating solar proton spectra. We calculate the spectrum based on flux measurements from spacecraft, which include a flux with energies between 100 and 500 MeV. In contrast, the calculations based on neutron monitors (e.g. [11,12]) largely ignores particles of these energies, focusing instead on particles with higher energies. Additionally, we assume that more solar protons reach the atmosphere during strong magnetic storms.

The results that different highlight the importance of model harmonisation that is currently work in progress by Space Weather Center Coordination Group. We expect the fruitful outcome that would increase the confidence among models used by the centers and ultimately increase safety for aviation. Along with harmonisation, the regular scientific-grade onboard dosimetry data will come handy for verification and tuning of the models. As SPEs cannot be forecasted the required equipment shall be installed onboard a small portion of operation civil aviation fleet in order to acquire dosimetry data in case future SPEs happen.

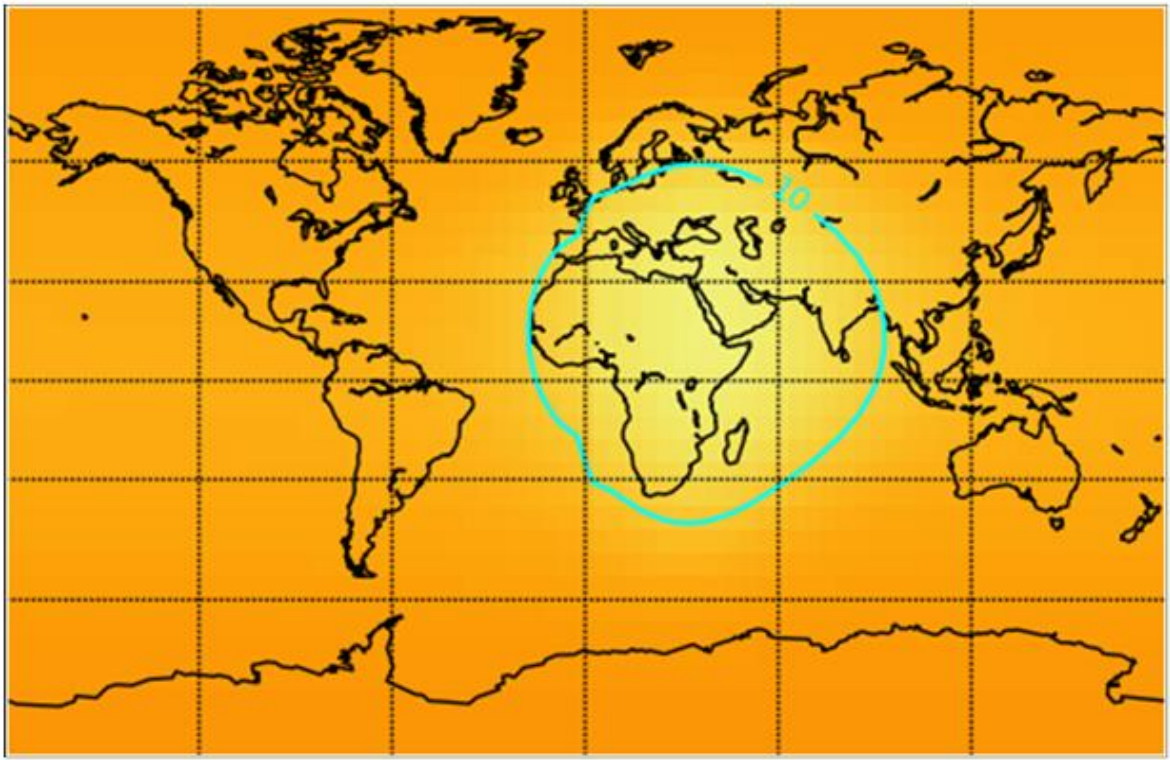

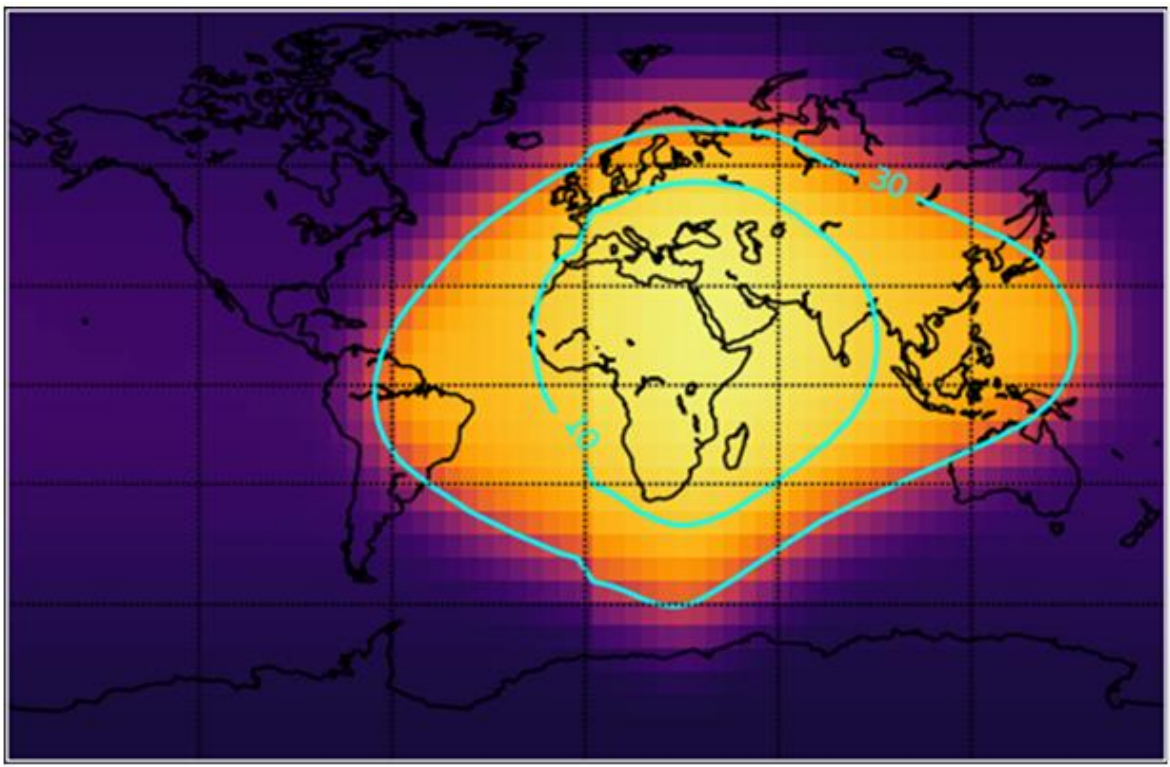

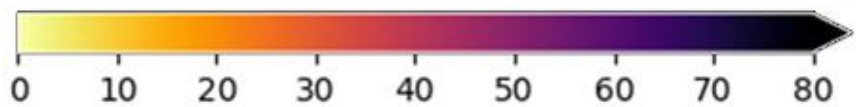

**Fig. 4**, panel 1 & 2: Radiation dose rate maps (in µSv/hour) for altitude of 12.2 km (40000 ft) during the onset of the event (02:08:22 UTC and 02:18:23 UTC).

**Acknowledgments** Data shown is provided by IAG, NOAA, GFZ, as aggregated at the CRC operations facility at Fedorov Institute of Applied Geophysics. This work is supported by State Assignments of Schmidt Institute of Physics of the Earth of the Russian Academy of Sciences and Geophysical Center of the Russian Academy of Sciences.